\documentclass[twocolumn,prb,showpacs,preprintnumbers,superscriptaddress,aps,amsmath,amssymb]{revtex4}
\usepackage{graphicx,color}

\begin{document}


\title{Spin Dynamics in $S=1/2$ Chains with Next-Nearest-Neighbor
Exchange Interactions}
\author{M.~Ozerov}
\affiliation{Dresden High Magnetic Field Laboratory (HLD),
Forschungszentrum Dresden-Rossendorf (FZD), 01314 Dresden, Germany}
\author{A.A.~Zvyagin}
\affiliation{{Institut f\"ur Festk\"orperphysik}, Technische Universit\"{a}t
Dresden, 01069 Dresden, Germany}
\affiliation{B.~Verkin Institute for Low
Temperature Physics and Engineering, National Academy of Science of Ukraine,
Kharkov 61103, Ukraine}
\author{E.~\v{C}i\v{z}m\'{a}r}
\affiliation{Dresden High Magnetic Field Laboratory (HLD),
Forschungszentrum Dresden-Rossendorf (FZD), 01314 Dresden, Germany}
\affiliation{Centre of Low Temperature Physics, P.J. \v{S}af\'{a}rik
  University, SK-041 54 Ko\v{s}ice, Slovakia}
\author{J.~Wosnitza}
\affiliation{Dresden High Magnetic Field Laboratory (HLD),
Forschungszentrum Dresden-Rossendorf (FZD), 01314 Dresden, Germany}
\affiliation{{Institut f\"ur Festk\"orperphysik}, Technische Universit\"{a}t
Dresden, 01069 Dresden, Germany}
\author{R.~Feyerherm}
\affiliation{Helmholtz-Zentrum Berlin (HZB) f\"{u}r Materialien und
Energie GmbH, 12489 Berlin, Germany}
\author{F.~Xiao}
\author{C.P.~Landee}
\affiliation{Department of Physics and Carlson School of Chemistry, Clark
University, Worcester, MA 01060, USA}
\author{S.A. Zvyagin}
\affiliation{Dresden High Magnetic Field Laboratory (HLD),
Forschungszentrum Dresden-Rossendorf (FZD), 01314 Dresden, Germany}

\date{\today}

\begin{abstract}
Low-energy magnetic excitations in the spin-1/2 chain compound (C$_6$H$_9$N$_2$)CuCl$_3$ [known as (6MAP)CuCl$_3$]
are probed by means of tunable-frequency electron spin resonance. Two modes with asymmetric (with respect to the $h\nu=g\mu_B
B$ line) frequency-field dependences are resolved, illuminating the striking incompatibility   with a simple uniform $S=\frac{1}{2}$ Heisenberg chain  model. The unusual ESR
spectrum is explained in terms of the recently developed theory for
spin-1/2 chains, suggesting the important role of
next-nearest-neighbor interactions in this compound.  Our conclusion is supported by
model calculations for the magnetic susceptibility of (6MAP)CuCl$_3$, revealing
a good qualitative agreement with experiment.
\end{abstract}


\maketitle
A spin-1/2 Heisenberg antiferromagnetic (AF) chain is one of the  paradigms in modern
quantum many-body physics. The most important feature of this model
is its integrability by means of the famous Bethe ansatz.
\cite{Bethe} However, as revealed theoretically and experimentally,
the uniform spin chains are unstable with respect to any
perturbation, breaking the chain uniformity. Such instability gives
rise to a rich variety of strongly correlated spin states and
quantum phase transitions, making these objects an attractive ground
for testing various theoretical concepts experimentally. A
competition between
nearest-neighbor (NN)  and next-nearest-neighbor (NNN) interactions,
as well as the presence of magnetic anisotropy, can fundamentally
modify the ground-state properties of quantum spin chains resulting
in a large diversity of complex magnetic structures.
\cite{Zb,Pinc,TM,Dr,ADM} The experimental determination of these
 interactions often is a challenging task. For that, magnetic  and
thermodynamic measurements (magnetization, magnetic susceptibility, and specific heat) are very helpful but do not give a detailed  picture of magnetic interactions.
Inelastic neutron scattering is a more suitable
tool, but it has certain serious limitations (including, for
instance, requirements on the sample size and chemical composition
of the material). That is why the search for new approaches (both
theoretical and experimental) that are helpful in clarifying the
microscopic structure of magnetic interactions appears to be of
particular importance.

Electron spin resonance (ESR) is traditionally recognized as one of
the most sensitive techniques for probing magnetic excitations in spin systems with collective ground states (see,
for instance, Refs.~\onlinecite{Hassan,Katsumata,Hill,Zvyagin_sol}).
Here, we present ESR studies of the low-energy excitation spectrum
in the $S=\frac{1}{2}$ chain system (C$_6$H$_9$N$_2$)CuCl$_3$
[hereafter (6MAP)CuCl$_3$]. Two gapped modes with asymmetric (with respect to the $h\nu=g\mu_B
B$ line) resonance positions
have been observed in the low-temperature ESR spectrum, reflecting the discrepancy with the simple spin-1/2 Heisenberg AF chain model,  employed for this compound previously. \cite{Geiser}   Our data are interpreted in the frame
of the recently developed mean-field-like theory for spin-1/2 chains, \cite{ZESR}
strongly suggesting   a  multiplication of the magnetic unit cell and  the presence of NNN
interactions in this compound.

\begin{figure}[b]
\begin{center}
\includegraphics[width=0.45\textwidth]{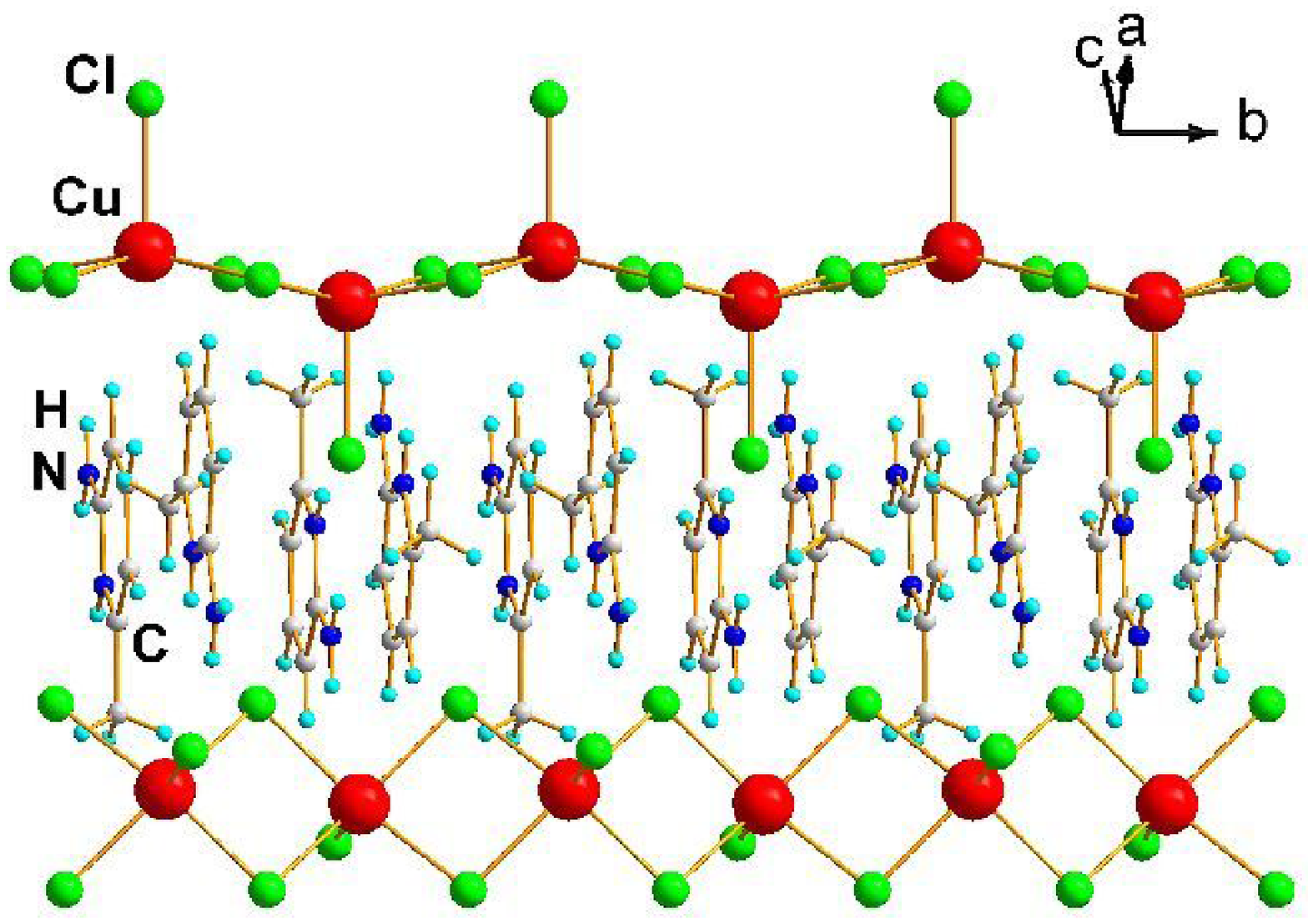}
\caption{\label{fig:6MAP-structure} (Color online) Crystal
structure of (6MAP)CuCl$_3$.}
\end{center}
\end{figure}

(6MAP)CuCl$_3$ crystallizes in the orthorhombic structure belonging
to the space group $Pnma$, with lattice constants $a=11.4$~{\AA},
$b=6.6$~{\AA}, and $c=12.8$~{\AA} (determined at room temperature). \cite{Geiser} The
number of formula units per unit cell is $Z=4$.  The
compound is built up from well-isolated, doubly bridged linear
chains of Cu$^{2+}$ ions (Fig.\ \ref{fig:6MAP-structure}). There are
two types of crystallographically similar chains running along the
$b$ axis. Each copper ion has a square-pyramidal coordination
geometry, with the axial bond substantially longer than the basal
ones, and with the direction of the axial bond alternating along the
chain axis $b$. The Cu-Cl-Cu bridging angle is well above
90$^{\circ}$, which defines the AF nature of the NN exchange
interactions [$J/k_B\approx 110$~K  \cite{Geiser,Liu}].  Using the formula for the N\'eel temperature
$T_N$ from Ref.~\onlinecite{Yasuda} and assuming $T_N < 100$ mK (as
evident from muon spin-relaxation experiments \cite{Blundell}), the
interchain interaction $J'/k_B$ is estimated to be less than 40~mK,
suggesting $J'/J<4\times10^{-4}$ and indicating an almost
perfect one-dimensional nature of the magnetic correlations in
(6MAP)CuCl$_3$. However, below about 15~K a deviation from the behavior
expected for a spin-1/2 uniform Heisenberg AF chain   appears, \cite{Geiser,Liu,Ozerov} i.e., the low-temperature magnetic susceptibility shows a pronounced low-temperature upturn. Very often, in spin-chain compounds such a
low-temperature tail in magnetic susceptibility originates from the
broken-chain effects and/or defects, overshadowing the intrinsic
low-temperature susceptibility  behavior (which  is of particular
importance when describing the ground state of spin chain
materials). To get a deeper insight into the  ground state  in (6MAP)CuCl$_3$ we decided to probe the
low-temperature excitation spectrum in this compound  by use of ESR
measurements.

Experiments have been  performed at the Dresden High Magnetic Field Laboratory (Hochfeld Magnetlabor Dresden, HLD) using  an X-band spectrometer (Bruker ELEXSYS E500) at a~fixed frequency of 9.4~GHz and  a~tunable-frequency ESR spectrometer (similar to that described in Ref.~\onlinecite{spectrometer}). High-quality single-crystalline
(6MAP)CuCl$_3$ samples with a typical size of 1.5x1.5x4 mm$^3$ were used. The magnetic field was applied along the $b$ axis. The magnetization
measurements have been performed by use of  ``Quantum Design'' Physical Property Measurement System (PPMS).

\begin{figure}[b]
\begin{center}
\includegraphics[width=0.95\columnwidth]{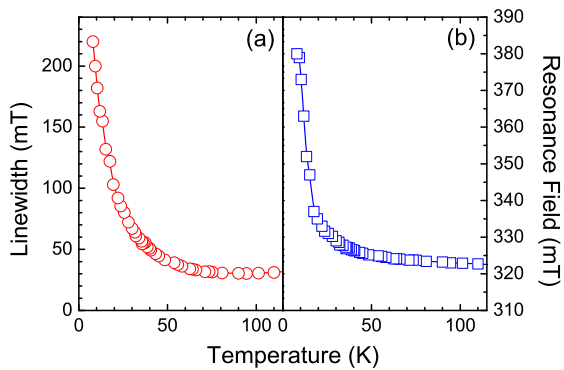}
\caption{\label{fig:LW} (Color online) Temperature dependence of the linewidth (a) and resonance field (b) measured at a frequency of 9.4  GHz. Lines are guides to the eye. }
\vspace{-1cm}
\end{center}
\end{figure}

A single ESR line (with $g=2.06$ measured at 20~K) was observed at
temperatures above 5~K. It was shown theoretically, \cite{OAZ} that  the ESR response of an ideal uniform Heisenberg spin chain with magnetically isotropic interactions should reveal a single peak in the ESR absorption with zero linewidth, \cite{how} and with the position of the resonance proportional to the applied  magnetic field. Spin-spin correlations would  yield a shift of the resonance position and a broadening of the ESR line {\em only} in the presence of  magnetic anisotropy. The temperature dependence of ESR linewidth and field  measured at frequency 9.4 GHz down to 7 K is shown in Fig.~\ref{fig:LW}, confirming unambiguously the  incompatibility  of the observed ESR behavior with the simple uniform  $S=\frac{1}{2}$ Heisenberg chain  model.
The low-temperature upturn of the linewidth, seen in Fig.~\ref{fig:LW}, can only be explained by the presence of relevant perturbations (from the viewpoint of renormalization group) of the critical Heisenberg chain [for instance, the  multiplication of unit cell, caused by  symmetric (exchange) or asymmetric (Dzyaloshinskii-Moriya) spin-spin interactions, or the $g$-tensor alternation]. \cite{OAZ}

\begin{figure}[t]
\begin{center}
\includegraphics[width=0.95\columnwidth]{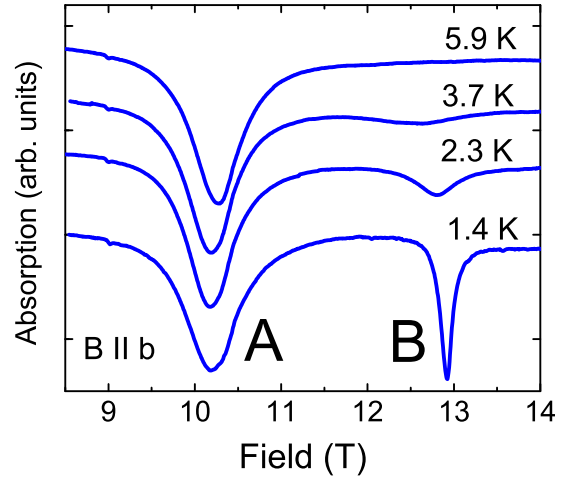}
\caption{\label{fig:6MAP-spectra} (Color online) ESR spectra of
(6MAP)CuCl$_3$ obtained at a frequency of 300 GHz
for different temperatures.}
\end{center}
\end{figure}

\begin{figure}
\begin{center}
\includegraphics[width=1\columnwidth]{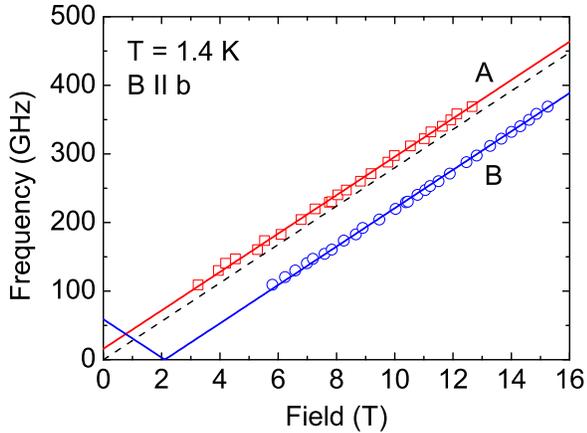}
\caption{\label{fig:6MAP-ffd} (Color online) Field dependence of
the magnetic-excitation frequencies in (6MAP)CuCl$_3$ at
$T=1.4$~K. Symbols denote experimental data, while solid lines correspond to
theoretical results (see text for details).  The dashed line denotes
$h\nu=g\mu_B B$ for  $g = 2$ (where $h$ is the Planck constant, $\nu$ is the
excitation frequency,  $\mu_B$ is Bohr's magneton, and
$B$ is the magnetic field).}
\end{center}
\end{figure}

To study the frequency-field dependence of the magnetic excitations in (6MAP)CuCl$_3$, further experiments were performed using the high-field (up to  16 T) tunable-frequency ESR spectrometer.
Below $\sim 5$~K, a second resonance mode (the mode B, Fig.~\ref{fig:6MAP-spectra}) appears. The frequency-field diagram of  the magnetic excitations  in (6MAP)CuCl$_3$ obtained at 1.4~K is
shown in Fig.~\ref{fig:6MAP-ffd}. The resonance absorptions in the
studied frequency-field range have a linear field dependence with
$g\approx2$. Linear extrapolations of the frequency-field
dependences of the ESR modes A and B to zero field yield the resonance-field shifts
$\Delta_A=15.9$~GHz (0.76~K) and $\Delta_B=-59$~GHz (-2.83~K),
respectively. The observation of two gapped modes clearly indicates the incompatibility  with the simple uniform magnetically isotropic spin-1/2 Heisenberg AF model,  suggesting the presence of additional interactions in this compound.

Recently, the dynamical mean-field-like
theory for the ESR in spin-1/2 chains with NN and NNN interactions (zigzag spin ladders)
has been developed. \cite{ZESR} It
was shown that in the case of an alternation of NN interactions, two ESR modes
at the frequencies $h\nu_{1,2} = |g\mu_B B +
\Delta_{1,2}|$ should emerge, where
\begin{eqnarray}
&&\Delta_{1,2} = {1\over 2}\biggl[(J_1+J_2+A_1+A_2)R_+ \nonumber \\
&&\pm \biggl( (J_1+J_2-2A_N)^2R_+^2 +(A_1+A_2-2A_N) \nonumber \\
&&\times (2J_1+2J_2 -2A_N+A_1+A_2)R_-^2\biggr)^{1/2}\biggr]
 \ . \
\end{eqnarray}
Here $B$ is the value of the applied magnetic  field, $\mu_B$ is  Bohr's magneton, $J_1$ and   $J_2$  are   spin-spin isotropic  interactions along the chain,
$A_1$ and $A_2$ denote the magnetic anisotropy,  while  $A_{N}$ is the  magnetically
anisotropic NNN interaction. $R_{\pm} = \langle S_{0,1}^z
\rangle \pm \langle S_{0,2}^z \rangle$ are the sum and the
difference of average values of the projections of the magnetic moments of
two magnetic centers (in the absence of the oscillating microwave field),  respectively.
Importantly, depending on the sign and strength of NN and NNN interactions, the theory  predicts different frequency-field diagrams of magnetic excitations which can be observed using ESR.  ESR experiments on the frustrated spin-1/2 quasi-one-dimensional systems In$_2$VO$_5$, \cite{Moller}  Li$_2$ZrCuO$_4$, \cite{Vavilova}
and on the asymmetric spin-ladder compound IPA-CuCl$_3$
\cite{Manaka} revealed qualitative agreement with the theoretical
predictions.

Contrary to uniform spin-1/2 Heisenberg AF chains  the combined effect of alternation and magnetic anisotropy  is predicted to manifest itself in the opening of a gap in the ESR spectrum. Most importantly,  the presence of  the AF NN and large enough NNN interactions  [$A_N [\langle S_{0,1}^z\rangle^2+\langle S_{0,2}^z\rangle^2]>(A_1+A_2)\langle S_{0,1}^z\rangle \langle S_{0,2}^z \rangle$]
 should result in  asymmetric (with respect to $h\nu=g\mu_B B$) frequency-field dependences of ESR peaks(Fig. 4 in Ref.~\onlinecite{ZESR}). This proposed frequency-field dependence  is  consistent with our observations in (6MAP)CuCl$_3$, strongly suggesting the presence of an NNN interaction and alternation in this compound.  Hydrogen position disorder in the amino groups linking  the Cu chains \cite{Geiser} can be regarded  as the origin of the  proposed alternation \cite{SP} (see discussion below).

\begin{figure}
\begin{center}
\includegraphics[width=1\columnwidth]{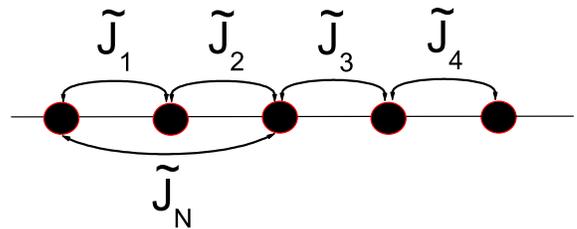}
\vspace{-2.5cm}
\caption{\label{fig:6MAP-chain}  A schematic view of a spin-chain structure with four
nearest-neighbor (\emph{\~{J}}$_{1,2,3,4}$)  and one next-nearest-neighbor (\emph{\~{J}}$_{N}$)
interactions proposed for (6MAP)CuCl$_3$ (see text for details). }
\end{center}
\end{figure}

The presence of additional interactions responsible for the energy-gap
opening in the low-energy ESR spectrum should manifest itself also
in a peculiar behavior of the magnetic susceptibility at temperatures
$T\sim \Delta_{1,2}/k_{B}$. On the other hand, the observation of a
broad maximum in the susceptibility of (6MAP)CuCl$_3$ at $T\sim 70$
K implies  the presence of short-range-order spin correlations at this
energy scale. Hence,  additional magnetic
eigenstates  with energies much
larger than the gaps, observed in the ESR experiments, need to be included.
To describe both the magnetic susceptibility and the ESR data, a minimal model has to include four magnetic sublattices.  Different values  of exchange couplings (defined as \emph{\~{J}}$_{1,2,3,4}$, Fig.~\ref{fig:6MAP-chain}) between neighboring sites along the chain would yield the multiplication
of the magnetic unit cell,   and  as a consequence
 four ESR modes in the excitation spectrum.

\begin{figure}
\begin{center}
\includegraphics[width=0.95\columnwidth]{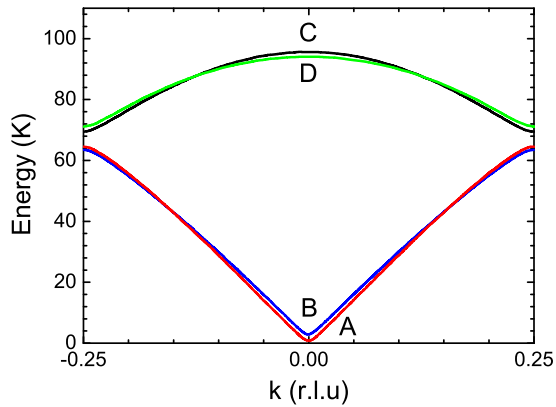}
\caption{\label{fig:disp4} (Color online) Dispersion relations of magnetic excitations at $B=0$,
determined by use of a four-center spin-chain  model.}
\end{center}
\end{figure}

\begin{figure}
\begin{center}
\includegraphics[width=0.95\columnwidth]{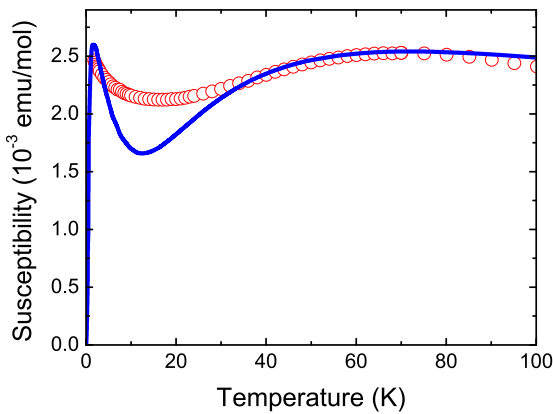}
\caption{\label{fig:chi} (Color online) Temperature dependence of
the magnetic susceptibility of (6MAP)CuCl$_3$ with a magnetic field of 0.1~T
applied along the $b$ axis. Experimental data are shown by symbols,
while the solid line corresponds to results of calculations for
the four-center model with the same set of parameters as in
Fig.~\ref{fig:disp4} (see the text for details).}
\end{center}
\end{figure}

Since no integrable four-center Heisenberg AF spin model for
(6MAP)CuCl$_3$ is available, the simplified four-center XY spin-chain model
\cite{Z1} is applied to illustrate the proposed scenario. The exceptional usefulness of this model  is determined by its  solvability.  The
dispersion laws of the four excitation branches, A, B, C, and D, are
shown in Fig.~\ref{fig:disp4}.
The best agreement between  the ESR data and calculations was
obtained  using the parameters \emph{ \~{J}}$_1/k_B = $\emph{\~{J}}$_3/k_B = 105$~K, \emph{\~{J}}$_2/k_B = 110$~K,
\emph{\~{J}}$_4/k_B = 96$~K, and \emph{\~{J}}$_{N}/k_B = 0.8$~K for the alternating NN (\emph{\~{J}}$_{1-4}$)
and NNN (\emph{\~{J}}$_{N}$) interactions, respectively.  The theory predicts also the existence of two  higher-energy ESR transitions (modes C and D).

In Fig.~\ref{fig:chi}, we show the temperature dependence of the magnetic susceptibility
(down to 1.8 K) together with the calculation results (based on the four-center chain model, Fig.~\ref{fig:6MAP-chain})  using the obtained parameters.
As follows from the model, the presence of low-energy
ESR gaps observed in our experiments should determine the low-$T$ part of
the magnetic susceptibility, while the high-energy branches (and
the related gaps) are responsible for the high-$T$ magnetic susceptibility dependence.
The main features of the
calculated susceptibility behavior (low-temperature maximum, a pronounced dip at $T\approx
13$~K,  and a broad maximum at $T\approx
70$~K) are consistent with the experimental data, suggesting the
validity of the four-center spin-chain model with NNN interaction. Detailed theoretical studies
of  the four-center AF spin chain with NN and NNN
interactions (unfortunately, not available now) are necessary to
precisely determine  relevant characteristics (exchange integrals, magnetic anisotropy, etc.) of the studied compound.

There are two previously known  compounds with an alternating spin-1/2 Heisenberg chain structure,  that are structurally uniform at room temperature: Cu(4-methylpyridine)$_2$Cl$_2$ [CuCl$_2$(mepy)$_2$]  \cite {Marsh81} and Cu(N-methylimidazole)$_2$ Br$_2$  [CuBr$_2$(midz)$_2$]. \cite {vanOoijen79}  The alternation parameter (ratio of weaker to stronger interactions strength was found to be 0.6(1) for CuCl$_2$(mepy)$_2$ and 0.4(1) for CuBr$_2$(midz)$_2$. Although  due to the tendency of these materials to twin in neither case  the methyl hydrogen ions could be located,  the uniform nature of the chains at room temperature is verified. Studies of the dielectric constants of the two compounds clearly revealed anomalies  near 50 K for CuCl$_2$(mepy)$_2$ \cite{Hall81} and at 50  and 105 K for CuBr$_2$(midz)$_2$ \cite{Wolthuis85}. These data were interpreted as  freezing of the methyl-group rotations which could induce slight twisting of the copper octahedra so as to produce an alternating variation of the exchange pathways. We are  not aware of any low-temperature structural study that confirms this hypothesis. Nevertheless, magnetic studies of these materials \cite{Smit79, deGroot82} revealed the characteristic  maximum in the magnetic susceptibility  as well as a gapped ground state [similar to that observed in (6MAP)CuCl$_3$].

While the methyl groups in (6MAP)CuCl$_3$ are ordered at room temperature, the hydrogen atoms of the amino groups are disordered \cite{Geiser}, leaving open the possibility that the observed magnetic behavior arises from a similar freezing transition. A potential signature of the proposed scenario  is a pronounced field-independent  maximum in the specific heat  at $\sim 2.3$ K \cite{Ozerov}, which would correspond to a low-temperature Schottky anomaly due to the hydrogen position order. Low-temperature structural studies of (6MAP)CuCl$_3$ are planned. It is notable that the variations of the exchange strengths in (6MAP)CuCl$_3$ are much smaller than those observed in CuCl$_2$(mepy)$_2$ and CuBr$_2$(midz)$_2$.

To summarize, employing the recently developed theory for ESR in
alternating spin chains the asymmetric double-peak ESR structure
observed in (6MAP)CuCl$_3$ is interpreted  as a signature of  NNN
interactions. The proposed four-center spin-chain model  consistently describes both the observed ESR spectrum and the
magnetic susceptibility.   Hence, based on the proposed model our
study has demonstrated the  potential feasibility of using ESR to probe NNN interactions
in alternating spin chains with  high resolution (available for ESR). Our approach   can be of
particular importance for understanding the nature of the ground-state and low-temperature magnetic properties of a wide range of
spin-chain systems with competing NN and NNN exchange interactions (including zig-zag spin ladders). Our observations call for systematic low-temperature structural and
neutron scattering studies of (6MAP)CuCl$_3$, which would allow  to verify the proposed model.

The authors would like to thank  M.M. Turnbull,  A.K. Kolezhuk, and M. Zhitomirsky for
fruitful discussions. This work was partly supported by the Deutsche
Forschungsgemeinschaft and EuroMagNET (under  Contract No. 228043).
E.\v{C}. appreciates the support under Grants No APVV-VVCE-0058-07 and
APVV-0006-07 from Slovak Research and Development Agency. A.A.Z. appreciates the
support from the Deutsche Forschungsgemeinschaft via the Mercator Program and  from the Institute of
Chemistry of the V. N.~Karazin Kharkov National University. We thank Tom Lancaster and Stephen Blundell (Oxford University) for performing the muon spin-relaxation experiment on (6MAP)CuCl$_3$.


\begin{thebibliography}{99}
\bibitem{SP} Such an alternation can be caused, for instance, by
different magnetic interactions along the chain as result of a
spin-Peierls transition. However, a thorough search for possible
superstructures by synchrotron x-ray diffraction, carried out at the
MAGS beamline at HZB, gave no indication for a
crystallographic/structural anomaly (which would suggest a
spin-Peierls transition) down to 1.5 K.
\end{thebibliography}
\end{document}